\documentclass[aps,preprintnumbers,showpacs,showkeys,nofootinbib,
superscriptaddress,fleqn,floatfix,tightenlines,preprint]{revtex4-1}
\usepackage{amsmath,amsfonts,amssymb,amscd,amsxtra,amsthm}
\usepackage{graphicx,slashed,bm}
\usepackage{mathrsfs,multirow,booktabs,epstopdf}

\usepackage[normalem]{ulem} 
\usepackage[dvipsnames]{xcolor} 
\renewcommand\sout{\bgroup \color{red} \ULdepth=-.5ex \ULset}

\begin{document}

\preprint{INHA-NTG-02/2015}
\title{Production of strange and charmed baryons in pion induced reactions}
\author{Sang-Ho Kim}
\email[E-mail: ]{shkim@rcnp.osaka-u.ac.jp}
\affiliation{Research Center for Nuclear Physics (RCNP), 
Osaka University, Ibaraki, Osaka, 567-0047, Japan}
\affiliation{Asia Pacific Center for Theoretical Physics (APCTP), Pohang,
Gyeongbuk, 790-784, Republic of Korea}
\author{Atsushi Hosaka}
\email[E-mail: ]{hosaka@rcnp.osaka-u.ac.jp}
\affiliation{Research Center for Nuclear Physics (RCNP), 
Osaka University, Ibaraki, Osaka, 567-0047, Japan}
\affiliation{J-PARC Branch, KEK Theory Center, 
Institute of Particle and Nuclear Studies, KEK, 
Tokai, Ibaraki, 319-1106, Japan}
\author{Hyun-Chul Kim}
\email[E-mail: ]{hchkim@inha.ac.kr}
\affiliation{Research Center for Nuclear Physics (RCNP), 
Osaka University, Ibaraki, Osaka, 567-0047, Japan}
\affiliation{Department of Physics, Inha University, Incheon 402-751,
Republic of Korea}
\affiliation{School of Physics, Korea Institute for Advanced Study
  (KIAS), Seoul 130-722, Republic of Korea}
\author{Hiroyuki Noumi}
\email[E-mail: ]{noumi@rcnp.osaka-u.ac.jp}
\affiliation{Research Center for Nuclear Physics (RCNP), 
Osaka University, Ibaraki, Osaka, 567-0047, Japan}
\date{\today}
\begin{abstract}
We study the strangeness and charm productions induced by the pion
beam, i.e. the $\pi^- p \to K^{*0}\Lambda$ and $\pi^- p \to
D^{*-}\Lambda_c^+$ reactions, based on two different theoretical
frameworks: an effective Lagrangian method and a Regge approach.  
In order to estimate the magnitude of the charm production relative to
that of the strangeness production, we assume that the coupling
constants for the charmed meson and baryon vertices are the same as
those for the strangeness channel. We found that the total cross
section for the charm production was about $10^3-10^6$ times smaller
than that for the strangeness production, depending on theoretical
approaches and kinematical regions. We also discuss each 
contribution to observables for both reactions. In general, the Regge
approach explains the experimental data very well over the whole
energy region. 
\end{abstract}
\pacs{13.75.Gx, 14.20.Jn, 14.20.Lq, 14.40.Lb}
\keywords{strange hadron productions, charmed hadron productions,
  effective Lagrangian, Regge approach}  

\maketitle
\section{Introduction} 
Charm-quark physics becomes one of the most important issues in
hadron physics, as experimental facilities report new 
hadrons containing one or two heavy quarks, either charm quarks or
bottom ones, with unprecedented precision. For example, the Belle
Collaboration, BABAR Collaboration, and BESIII Collaboration have
announced new
mesons~\cite{Choi:2003ue,Aubert:2004ns,Aubert:2005rm,Abe:2007jna, 
Choi:2007wga,Belle:2011aa,Ablikim:2013mio,Liu:2013dau,Ablikim:2013wzq}, 
some of which were also confirmed by the LHCb
Collaboration~\cite{Aaij:2013zoa,Aaij:2014jqa} (see
Refs.~\cite{Swanson:2006st,Brambilla:2010cs} for reviews). While the
mesons with charm have been extensively studied theoretically as
well as experimentally, charmed baryons have been investigated less often.
However, the charmed baryons are equally or even more
important, since they provide a good opportunity to examine the role
of both chiral symmetry and heavy quark symmetry in heavy-light quark
systems. Moreover, the structure and the production mechanism of the
charmed baryons are much less known than those of light-quark baryons.   
Recently, a new proposal was submitted at the J-PARC (Japan Proton
Accelerator Research Complex) facility for the study  
of charmed baryons via the pion induced reactions at a high-momentum 
beam line~\cite{e50}. There has been only one earlier work at
Brookhaven National Laboratory (BNL) almost 30 years ago to search  
for the charm productions associated with the mechanism $\pi^- p \to
D^{*-}B_c$, where $B_c$ denotes a charmed baryon in ground or excited 
states~\cite{Christenson:1985ms}. 

In Ref.~\cite{Kim:2014qha}, the differential cross sections
$d\sigma/dt$ for the strangeness and charm productions, i.e.  
$\pi^- p \to K^{*0}\Lambda$ and $\pi^- p \to D^{*-}\Lambda_c^+$, 
were investigated by using a simple Regge model. As expected, the
differential cross section for the charm production was found to be
much less than that for the strangeness production.  
In the present work, we want to further elaborate the previous
investigation of these two processes, employing both an effective
Lagrangian method and a Regge approach, while putting emphasis on the
latter. The effective Lagrangian method has been successfully used to
study the production of strangeness hadrons. There are two important
ingredients in this method: coupling constants and form
factors. The coupling constants are easily determined by using
well-known baryon-baryon potentials such as the Nijmegen potential or
by considering the experimental data for hadron decays. 
However, the cutoff masses of the form factors cause ambiguity in
describing reactions. In the Regge approach, we also have parameters
to fix.  In order to determine them, we utilize the quark-gluon string
model (QGSM) introduced by Kaidalov 
{\it et al}~\cite{Kaidalov:1982bq,Boreskov:1983bu,Kaidalov:1986zs,   
Kaidalov:1994mda}. In particular, Ref.~\cite{Boreskov:1983bu} studied
the $\pi^- p \to D^-\Lambda_c^+$ reaction within this model, relying
only on the $D^*$ reggeon. However, in this work, we consider the
contributions of the $D$ and $\Sigma_c$ reggeons as well as of the $D^*$ 
reggeon. 
Before we proceed, we want to shortly mention why we first
start with the $D^*\Lambda_c$ production rather than the $D\Lambda_c$
one. The reason lies in the fact that there exists a technical problem 
in experiment: the background can be reduced more efficiently 
in reconstructing $D^{*-}$ than in $D^-$. The decay  chain
$D^{*-} \to D^0 + \pi^- \to K^+ + \pi^- + \pi^-$ allows one to reduce
a combinatorial background more effectively.

The present work is organized as follows: In Sec. II, we briefly
explain the formalism of the effective Lagrangian method and then
present the numerical results of the total cross sections and
differential cross sections for pion induced $K^{*0}\Lambda$ and $  
D^{*-}\Lambda_c^+$ productions. In  
Sec. III, we derive the transition amplitudes within the Regge
approach. We also discuss in this section the results for both the
$\pi^- p \to K^{*0}\Lambda$ and $\pi^- p \to D^{*-}\Lambda_c^+$
reactions, based on the Regge approach. 
In Sec. IV, we compare the results from the Regge approach
with those from the effective Lagrangian method. We discuss the
comparison in detail. 
The final section is devoted to the summary and the conclusion of the 
present work. 

\section{Effective Lagrangian Approach}
In this section, we employ an effective Lagrangian approach to study
both the $\pi^- p \to K^{*0}\Lambda$ and $\pi^- p \to D^{*-}\Lambda_c^+$
reactions. Starting from the effective Lagrangians describing the
interactions between hadrons, we are able to construct the diagrams of
$t$-, $s$-, and $u$-channels at the tree level. This effective
Lagrangian method has been known to be successful in describing hadron
productions near threshold.  

\subsection{Lagrangians and Feynman amplitudes}
\begin{figure}[h]
\centering
\includegraphics[scale=0.6]{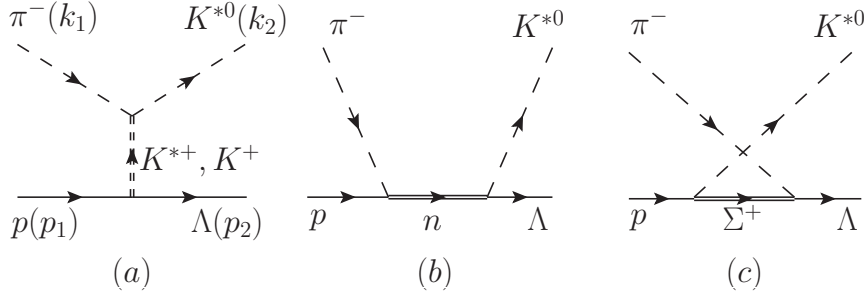}
\caption{Tree-level diagrams for $\pi^- p \to K^{*0} \Lambda$.}        
\label{FIG1}
\end{figure}
We first begin with the $\pi^- p \to K^{*0}\Lambda$ reaction.
The relevant Feynman diagrams are displayed in Fig.~\ref{FIG1} in
which $k_1$ and $p_1$ denote the initial momenta of the $\pi$ and the
proton, respectively. $k_2$ and $p_2$ stand for those of the final
$K^*$ and $\Lambda$, respectively. In this model, we include (a) $K$
and $K^*$ exchanges in the $t$-channel; (b) the nucleon exchange in
the $s$-channel; and (c) the hyperon ($\Sigma$) exchange in the
$u$-channel.

To obtain the invariant Feynman amplitudes, we use the following
Lagrangians:
\begin{eqnarray}
\mathcal L_{\pi K K^*} &=& 
-ig_{\pi K K^*} ( \bar K \partial^\mu \bm{\tau}\cdot\bm{\pi} K^*_\mu - 
               \bar K_\mu^* \partial^\mu \bm{\tau}\cdot\bm{\pi} K ),    \cr
\mathcal L_{\pi K^* K^*} &=&
g_{\pi K^* K^*} \varepsilon^{\mu\nu\alpha\beta}
\partial_\mu \bar K_\nu^* \bm{\tau}\cdot\bm{\pi} \partial_\alpha K_\beta^*,
\label{eq:Lag1}
\end{eqnarray}
for the $K^*$ meson and pseudoscalar-octet-meson interactions, 
where $\pi$, $K$, and $K^*$ designate the fields of $\pi(140,0^-)$, 
$K(494,0^-)$, and $K^*(892,1^-)$ mesons, respectively.
The coupling constant $g_{\pi K K^*}$ is determined by the
experimental data of the decay width $\Gamma(K^* \to K
\pi)$~\cite{Olive:2014zz}. The decay width is 
expressed in terms of the coupling constant 
\begin{equation}
\label{eq:DWidth}
\Gamma(K^* \to K \pi) = g_{K^* K \pi}^2 \frac{k_\pi^3}{8 \pi M_{K^*}^2},  
\end{equation}
where $k_\pi$ is the three-momentum of the decaying particle 
\begin{equation}
\label{eq:DMomentum}
k_\pi = 
\frac{\sqrt{[M_{K^*}^2-(M_K+M_\pi)^2][M_{K^*}^2-(M_K-M_\pi)^2]}}{2M_{K^*}}, 
\end{equation}
so that one finds $g_{\pi K K^*} = 6.56$. 
To determine the $\pi K^* K^*$ coupling constant $g_{\pi K^* K^*}$, we
use the hidden local gauge symmetry~\cite{Bando:1987br} and flavor
SU(3) symmetry. The value of the $\pi K^* K^*$ coupling constant is
obtained as $g_{\pi K^* K^*}=7.45\,\mathrm{GeV}^{-1}$. 

The effective Lagrangians for the pseudoscalar meson and  baryon octet
vertices are written as   
\begin{eqnarray}
\mathcal L_{\pi N N} &=&
\frac{g_{\pi N N}}{2M_N} \bar N \gamma_\mu
\gamma_5 \partial^\mu \bm{\tau}\cdot\bm{\pi} N,                       \cr
\mathcal L_{\pi \Sigma \Lambda} &=&
\frac{g_{\pi \Sigma \Lambda}}{M_\Lambda+M_\Sigma} \bar \Lambda \gamma_\mu
\gamma_5 \partial^\mu \bm{\pi}\cdot\bm{\Sigma} + \mathrm{H.c.},       \cr
\mathcal L_{K N \Lambda} &=&
\frac{g_{K N \Lambda}}{M_N+M_\Lambda} \bar N \gamma_\mu  
\gamma_5 \Lambda \partial^\mu K + \mathrm{H.c.}.       
\label{eq:Lag2}
\end{eqnarray}
where $N$, $\Lambda$, and $\Sigma$, denote the nucleon,
$\Lambda(1116)$, and $\Sigma(1190)$ baryon fields, respectively. The
values of the coupling constants $g_{\pi N N}=13.3$, $g_{\pi \Sigma
  \Lambda}= 11.9$, and $g_{K N \Lambda}=-13.4$ are taken from the
Nijmegen soft-core model (NSC97a)~\cite{Stoks:1999bz}. 

The interaction between the $K^*$ vector meson and the baryon octet
is described by the following effective Lagrangian
\begin{eqnarray}
\mathcal L_{K^* N Y} =
-g_{K^* N Y} \bar N \left[ \gamma_\mu Y - 
\frac{\kappa_{K^* N Y}}{M_N+M_Y} \sigma_{\mu\nu} Y 
\partial^\nu \right] K^{*\mu} + \mathrm{H.c.},   
\label{eq:Lag3}
\end{eqnarray}
where $Y$ generically stands for $\Lambda$ or
$\bm{\tau}\cdot\bm{\Sigma}$. We again take the values of the coupling
constants $g_{K^*NY}$ and $\kappa_{K^*NY}$ from the Nijmegen
potential~\cite{Stoks:1999bz} 
\begin{eqnarray}
g_{K^* N \Lambda} &=& -4.26, \,\,\, \kappa_{K^* N \Lambda} = 2.91,           \cr
g_{K^* N \Sigma} &=& -2.46, \,\,\, \kappa_{K^* N \Sigma} = -0.529.
\label{eq:Coupl}
\end{eqnarray}

The scattering amplitude for the $\pi N \to K^* \Lambda$ process can be
written as
\begin{eqnarray}
\mathcal M =
\varepsilon_\mu^* \bar{u}_\Lambda \, \mathcal M^\mu \,\, u_N,
\label{eq:NotatAmpl}
\end{eqnarray}
where $u_N$ and $u_\Lambda$ denote the Dirac spinors for the incoming 
nucleon and for the outgoing $\Lambda$, respectively, and  
$\varepsilon_\mu$ stands for the polarization vector of the outgoing
$K^*$ meson. The corresponding amplitude to each channel is obtained
as follows: 
\begin{eqnarray}
\mathcal M_{K}^\mu &=& I_K
\frac{ig_{\pi K K^*}}{t-M_K^2}
\frac{g_{K N \Lambda}}{M_N+M_\Lambda}
\gamma^\nu \gamma_5 k_1^\mu (k_2-k_1)_\nu,                               \cr
\mathcal M_{K^*}^\mu &=& I_{K^*}
\frac{g_{\pi K^* K^*} g_{K^* N \Lambda}}{t-M_{K^*}^2}
\epsilon^{\mu\nu\alpha\beta} 
\left [ \gamma_\nu - \frac{i\kappa_{K^* N \Lambda}}{M_N+M_\Lambda} 
\sigma_{\nu \lambda}(k_2-k_1)^\lambda \right ] k_{2 \alpha} k_{1 \beta},         \cr 
\mathcal M_{N}^\mu &=& I_N
\frac{i g_{K^* N \Lambda}}{s-M_N^2}
\frac{g_{\pi N N}}{2M_N}
\left [ \gamma^\mu - \frac{i\kappa_{K^* N \Lambda}}{M_N+M_\Lambda}
        \sigma^{\mu\nu} k_{2 \nu} \right ] 
(\rlap{/}{k_1}+\rlap{/}{p_1}+M_N)
\gamma^\alpha \gamma_5 k_{1\alpha},                                       \cr 
\mathcal M_{\Sigma}^\mu &=& I_\Sigma
\frac{i g_{K^* N \Sigma}}{u-M_{\Sigma}^2}  
\frac{g_{\pi \Sigma \Lambda}}{M_\Sigma+M_\Lambda}
\gamma^\alpha \gamma_5 (\rlap{/}{p_2}-\rlap{/}{k_1}+M_{\Sigma}) 
\left [ \gamma^\mu - \frac{i\kappa_{K^* N \Sigma}}{M_N+M_{\Sigma}}
        \sigma^{\mu\nu} k_{2 \nu} \right ] k_{1\alpha}.
\label{eq:EachAmpl}                           
\end{eqnarray}
The isospin factors are given as $I_K = I_{K^*} = I_N = I_\Sigma =
\sqrt{2}$. With the form factors taken into account, the total result
for the invariant amplitude is written as
\begin{eqnarray}
\mathcal M (\pi^- p \to K^{*0}\Lambda) =
\mathcal M_K \cdot F_K + \mathcal M_{K^*} \cdot F_{K^*}             
+ \mathcal M_N \cdot F_N + \mathcal M_\Sigma \cdot F_\Sigma.
\label{eq:SumAlpl}
\end{eqnarray} 
We choose the following form of the form factors
\begin{eqnarray}
F_{ex}(p^2) = \frac{\Lambda^4}{\Lambda^4+(p^2-M_{ex}^2)^2},
\label{eq:FF1}
\end{eqnarray}
where $p$ and $M_{ex}$ designate generically the transfer
momentum and the mass of the exchanged particle, respectively.
The cutoff mass $\Lambda$ is usually fitted to reproduce
the experimental data and depends on the reaction channel,
$K$, $K^*$, $N$, and $\Sigma$-exchanges. However, to minimize the number
of parameters for a rough estimation of the production rate, we employ
two different cutoff parameters for the meson exchanges and baryon
exchanges, separately, which are chosen to be 
$\Lambda_{K,K^*}=0.55\,\mathrm{GeV}$ and $\Lambda_{N,\Sigma}=0.60\,
\mathrm{GeV}$. 

\begin{figure}[h]
\centering
\includegraphics[scale=0.6]{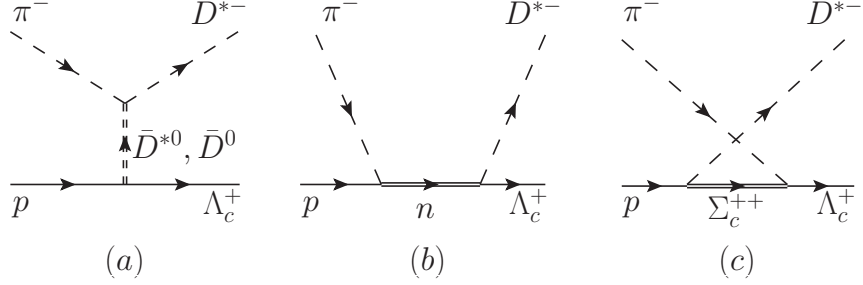}
\caption{Tree-level diagrams for $\pi^- p \to D^{*-}\Lambda_c^+$.}        
\label{FIG2}
\end{figure}

The Feynman amplitude $\mathcal M$ is related to the differential 
cross section as
\begin{eqnarray}
\frac{d\sigma}{dt} 
= \frac{1}{64\pi (p_{\mathrm{cm}})^2 s}
\frac{1}{2}\sum_{s_i,s_f,\lambda_f}|\mathcal M|^2,
\label{eq:Def:dsdt}
\end{eqnarray}
where $s_i$ and $s_f$ stand for the spins of the nucleon and the
$\Lambda$, respectively. $\lambda_f$ denotes the polarization label of
the $K^*$ meson and $p_{\mathrm{cm}}$ the momentum of the pion in the
center-of-mass frame.

Now we turn to the charm production reaction $\pi^- p \to 
D^{*-} \Lambda_c^+$. The relevant Feynman diagrams are depicted in
Fig.~\ref{FIG2}. The amplitude for this charm production reaction is
obtained just by replacing the strange mesons and hyperons with the 
charmed ones. In principle, the coupling constants for the charmed
hadrons should be different from those for the strange  
hadrons.
In the present calculation, however, we use the same strengths for the
corresponding vertices when the coupling constants are dimensionless.
This might be considered to be a good assumption if strange and charm 
quarks are sufficiently heavy. 
On the other hand, for the coupling constant $g_{\pi K^* K^*}$ which carries 
the dimension of the inverse mass, we introduce the scaling as
$g_{\pi D^* D^*} = M_{K^*} / M_{D^*} \cdot  g_{\pi K^* K^*}$.
In practice, it is known that the coupling constant $g_{\pi D D*}$ is about
twice as large as the $g_{\pi K K^*}$. This difference of the
strengths between $g_{\pi D D^*}$ and  $g_{\pi K K^*}$ could be the 
source of the ambiguity in the present calculation, which would
influence the magnitude of amplitudes. As for the form factors, we 
will use the same form as Eq.~(\ref{eq:FF1}) with the equal cutoff
masses. By doing that, we can directly compare the magnitude of the
total cross section for the $\pi N \to D^*\Lambda_c$ reaction with
that for the $\pi N \to K^*\Lambda$.  

\subsection{Results for $K^{*0}\Lambda$ production}
\begin{figure}[thp]
\centering
\includegraphics[scale=0.4]{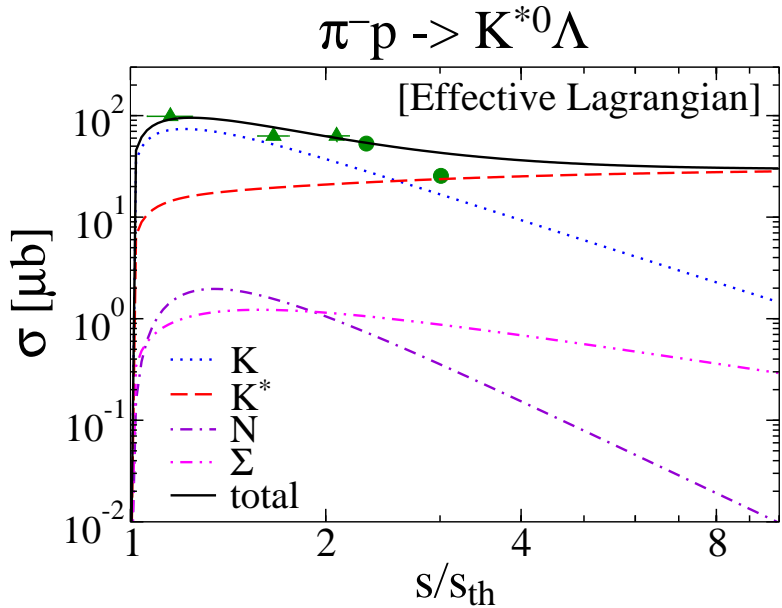}
\caption{(Color online). 
Each contribution to the total cross sections 
for the $\pi^- p \to K^{*0}\Lambda$ reaction given
as a function of $s/s_{\mathrm{th}}$,
based on an effective Lagrangian approach. 
The dotted and dashed curves show the contributions of 
$K$ exchange and $K^*$ exchange, respectively. 
The dot-dashed and dot-dot-dashed ones draw 
the effects of $N$ and $\Sigma$ exchanges, respectively.
The solid curve represents the total result. 
The experimental data are taken from Ref.~\cite{Dahl:1967pg} 
(triangles) and from Ref.~\cite{Crennell:1972km} (circles).}          
\label{FIG3}
\end{figure}

Let us first show contributions of each channel to the total cross
section for the reaction $\pi^- p \to K^{*0} \Lambda$. In
Fig.~\ref{FIG3}, they are drawn as a function of $s/s_{\mathrm{th}}$,
where $s_{\mathrm{th}}$ is the value of $s$ at threshold,
i.e. $s_{\mathrm{th}} = (m_{K^*}+m_\Lambda)^2 = 4.05 \,\,\,
\mathrm{GeV}^2$.  As shown in Fig.~\ref{FIG3}, the $t$-channel
process makes the most dominant contribution 
to the total cross section. $K$ exchange plays a crucial
role in describing the total cross section in the low-energy region,
whereas $K^*$ exchange governs its behavior in the high-energy region.    
The reason lies in the fact that the contribution of $K$ exchange
decreases as $s$ increases, while the effect of $K^*$ exchange becomes
larger than that of $K$ exchange as the value of $s/s_{\mathrm{th}}$
bocomes greater than around 3. Though the contribution of $K^*$ exchange
seems to increase as $s$ increases, it turns out to be almost constant
as $s$ becomes very large. In fact, one can show analytically that
when $s$ is very large, the total cross section is proportional to
$s^{J-1}$, where $J$ stands for the spin of an exchange particle in
the $t$-channel. This is the reason why $K$ exchange contributes
mainly to the low-energy region, whereas $K^*$ exchange comes into
play when $s/s_{\mathrm{th}}$ gets large. On the other hand, the
contribution of baryon exchanges is almost negligible over the whole
energy region. 

The result of the total cross section is in good agreement with the
experimental data~\cite{Dahl:1967pg,Crennell:1972km} in the relatively    
low-energy region ($s/s_{\mathrm{th}} \lesssim 2.1$). However, 
the present result starts to deviate from the experimental data when
$s/s_{\mathrm{th}}$ reaches the value of around 2.5. 
Generally, the effective Lagrangian method for the Born approximation at
the tree level is a good approximation for the lower-energy regions near
threshold, which, however, may not be used at high energies as it often
violates the unitarity.

\vspace{0.3cm}
\begin{figure}[thp]
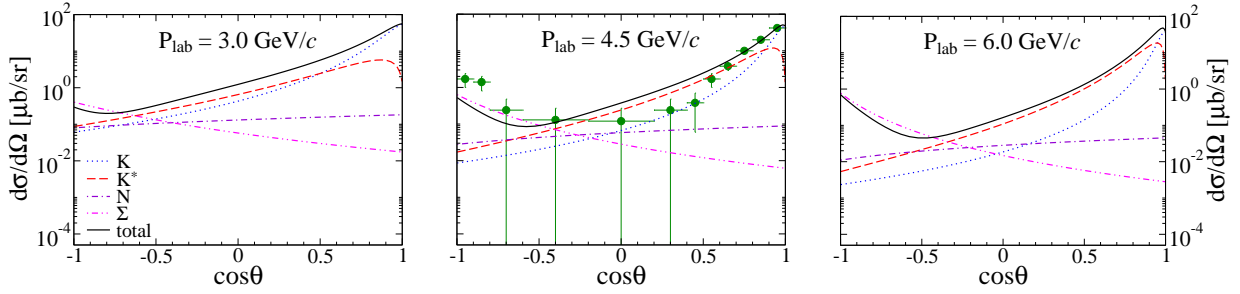

\centering
\includegraphics[width=5.30cm]{FIG4a.eps}\hspace{1.1em}
\includegraphics[width=4.55cm]{FIG4b.eps}\hspace{1.1em}
\includegraphics[width=5.30cm]{FIG4c.eps}
\caption{(Color online).
Differential cross sections for the $\pi^- p \to K^{*0}\Lambda$ reaction 
as functions of $\cos\theta$
at three different pion momenta ($\mathrm{P_{lab}}$), 
based on an effective Lagrangian approach.
The experimental data are taken from Ref.~\cite{Crennell:1972km}.
The notations are the same as Fig.~\ref{FIG3}.}
\label{FIG4}
\end{figure}

Each contribution to the differential cross section
$d\sigma/d\Omega$ for the $\pi^- p \to K^{*0}\Lambda$ reaction is 
illustrated in Fig.~\ref{FIG4} as a function of $\cos\theta$ at three
different momenta, i.e. $\mathrm{P_{lab}} = 3.0\,\mathrm{GeV}/c$, 
$4.5\,\mathrm{GeV}/c$, and $6.0\,\mathrm{GeV}/c$. Note that the
experimental data exist only for $\mathrm{P_{lab}}
=4.5\,\mathrm{GeV}/c$. The $\theta$ is the scattering angle between  
the incoming $\pi$ and the outgoing $K^*$ meson in the center-of-mass 
frame. As shown in Fig.~\ref{FIG4}, $K$ and $K^*$ exchanges make
similar contributions to $d\sigma/d\Omega$: Their effects diminish as
$\cos\theta$ decreases except that the contribution of $K^*$ exchange
is sharply reduced at the very forward angle. While $N$ exchange makes 
only a minor contribution, $\Sigma$ exchange in the $u$-channel becomes
dominant at the very backward angles. As $\mathrm{P_{lab}}$
increases, the $K$-exchange contribution diminishes faster with
$\cos\theta$ decreased. On the other hand, the $u$-channel
contribution reveals behavior opposite to the $t$-channel ones. 
Because of these different characters of each contribution, the dip
structure appearing in the differential cross section becomes deeper
as $\mathrm{P_{lab}}$ increases. 

The $t$-channel contribution explains the experimental data
~\cite{Crennell:1972km} very well in the forward direction at
$\mathrm{P_{lab}} = 4.5\, \mathrm{GeV}/c$, while they start to deviate
from the data, as $\cos\theta$ decreases. The result is underestimated
at the backward angles. Moreover, if one takes a close look at the
experimental data, one finds that $d\sigma/d\Omega$ turns flat between 
$\cos\theta=0.3$ and $\cos\theta=-0.7$. The dip structure
that the effective Lagrangian method produces is not enough to describe 
this flatness at intermediate angles. We will discuss this structure
in more detail later within the Regge approach.

\begin{figure}[htp]
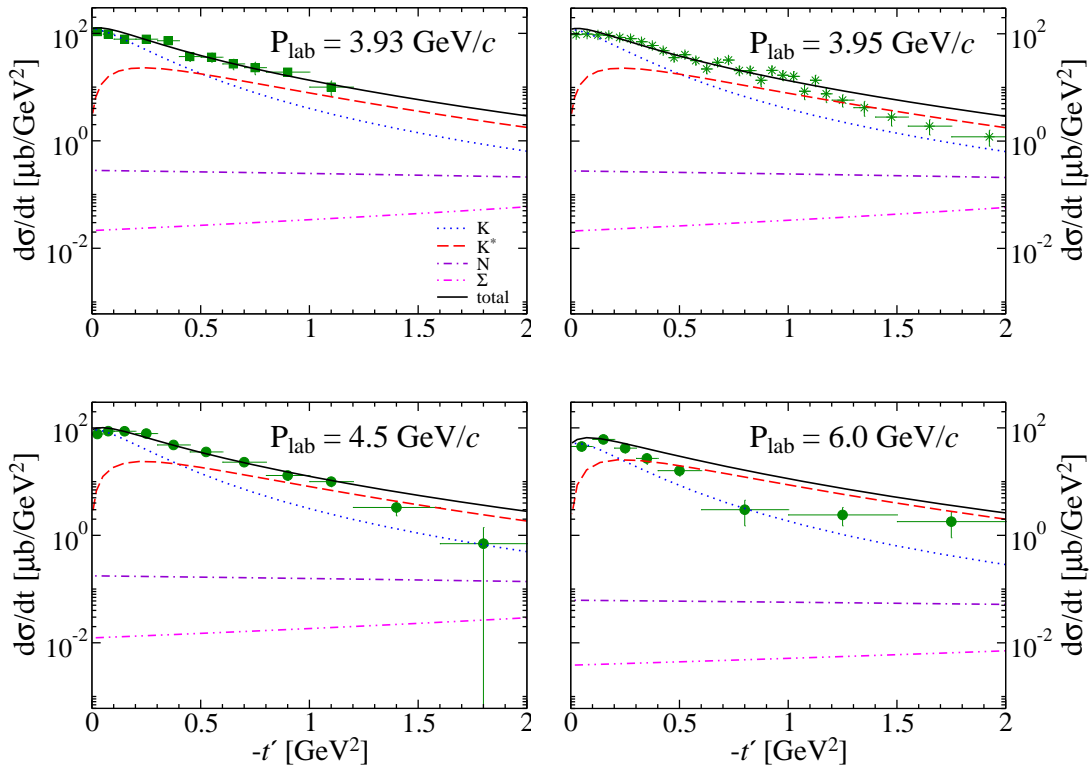

\vspace{1em}
\begin{tabular}{cc}
\includegraphics[width=7.0cm]{FIG5a.eps}\,\,\,\,
\includegraphics[width=7.0cm]{FIG5b.eps}\vspace{1.73em} \\ 
\includegraphics[width=7.0cm]{FIG5c.eps}\,\,\,\,
\includegraphics[width=7.0cm]{FIG5d.eps}
\end{tabular}
\caption{(Color online).
Differential cross sections for the $\pi^- p \to K^{*0}\Lambda$ reaction
as functions of $-t^\prime$ at four different pion momenta 
($\mathrm{P_{lab}}$), based on an effective Lagrangian approach.
The experimental data denoted by the squares are taken from
Ref.~\cite{Yaffe:1973ex}, 
and those denoted by the stars from Ref.~\cite{Aguilar-Benitez:1980pd}. 
Those designated by the circles are taken from 
Ref.~\cite{Crennell:1972km}. 
The notations are the same as Fig.~\ref{FIG3}.}
\label{FIG5}
\end{figure}
Figure~\ref{FIG5} shows the results of the differential cross
sections $d\sigma/dt$ for the $\pi^- p \to K^{*0}\Lambda$ reaction
at four different momenta $\mathrm{P_{lab}} = 3.93\,\mathrm{GeV}/c$,
$3.95\,\mathrm{GeV}/c$, $4.5\,\,\mathrm{GeV}/c$,  
and $6.0\,\mathrm{GeV}/c$, compared with the experimental data.
They are drawn as functions of $-t^\prime=t_{\mathrm{max}}-t$, where
the minimum and maximum values of $t$ are given kinematically as
\begin{eqnarray}
&&t_{\mathrm{min}}^{\mathrm{max}} = M_{\pi}^2 + M_{K^*}^2 - \frac{1}{2s}
\biggl[ 
[s - (M_N^2-M_\pi^2)][s - (M_\Lambda^2-M_{K^*}^2)]                        \cr
&&\mp
\sqrt{[s - (M_N+M_\pi)^2][s - (M_N-M_\pi)^2]}           %
 \cr  &&
\sqrt{[s - (M_\Lambda+M_{K^*})^2][s - (M_\Lambda-M_{K^*})^2]}       
\biggr],
\label{eq:tMaxMin}
\end{eqnarray}
respectively. For each of the fixed energies,
$t$ varies between $t_{\mathrm{min}}$ and $t_{\mathrm{max}}$ (or $-t'$
varies between 0 and $t_{\mathrm{max}} - t_{\mathrm{min}}$). 
The contributions of the $t$-channel decrease as
$-t'$ increases, as expected. $K$ exchange governs 
$d\sigma/dt$ near $-t'\approx 0$, whereas $K^*$ exchange
becomes the main contribution to $d\sigma/dt$. This
feature does not change in general, even though $\mathrm{P_{lab}}$ 
increases. The $s$- and $u$-channels are negligible. The results from the 
effective Lagrangian approach are in good agreement with the experimental 
data between $-t'=0$ and $-t'=1.2\,\mathrm{GeV}^2$, but start to deviate 
from the data as $-t'$ increases. Note that the effective Lagrangian 
method can only explain the data in the smaller $-t'$ region when
$\mathrm{P_{lab}}=6.0\,\mathrm{GeV}/c$. 

\subsection{Results for $D^{*-}\Lambda_c^+$  production}
\begin{figure}[thp]
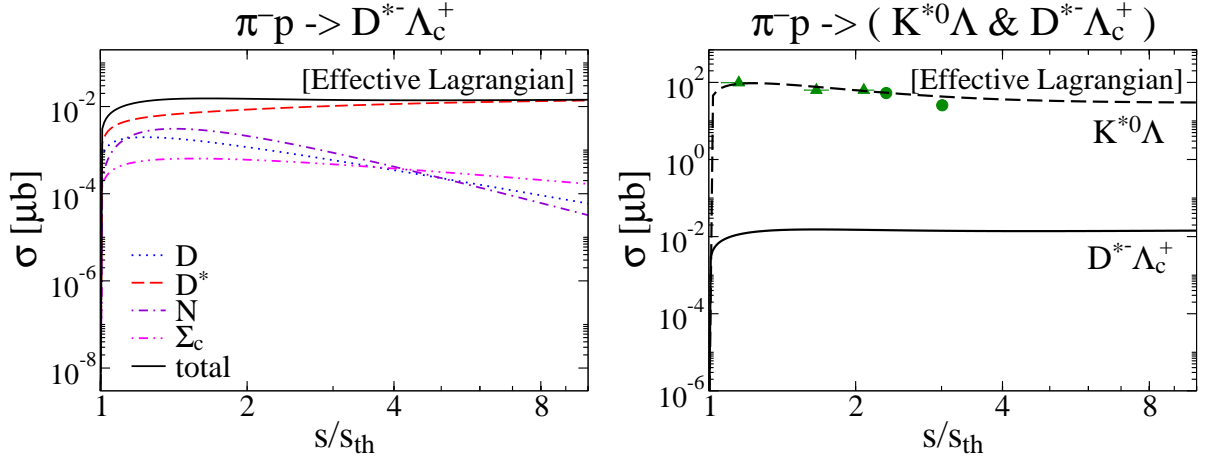

\begin{tabular}{cc} 
\includegraphics[scale=0.3]{FIG6a.eps} \;
\includegraphics[scale=0.3]{FIG6b.eps}
\end{tabular}
\caption{(Color online). 
In the left panel, each contribution to the total cross sections 
for the $\pi^- p \to D^{*-}\Lambda_c^+$ reaction is drawn 
as a function of $s/s_{\mathrm{th}}$ 
from an effective Lagrangian approach. 
The dotted and dashed curves show $t$-channel contributions, i.e. 
those of $D$ exchange and $D^*$ exchange, respectively. 
The dot-dashed and dot-dot-dashed ones depict 
the contributions of baryon exchange ($N$ and $\Sigma_c$), respectively. 
The solid curve represents the full result of the total cross section.
In the right panel, the total cross section for 
the $\pi^- p \to D^{*-}\Lambda_c^+$ reaction (solid curve) is compared with 
that for the $\pi^- p \to K^{*0}\Lambda$ one (dashed one). 
The experimental data for the $\pi^- p \to K^{*0}\Lambda$ reaction are 
taken from Ref.~\cite{Dahl:1967pg} (triangles) and 
from Ref.~\cite{Crennell:1972km} (circles).}       
\label{FIG6}
\end{figure}

We now turn to the charm production. In the left panel of
Fig.~\ref{FIG6}, the results of the total cross section as well as
various contributions for the $\pi^- p \to D^{*-}\Lambda_c^+$ reaction
are drawn as a function of $s/s_{\mathrm{th}}$. Note that
$s_{\mathrm{th}}$ is different from the case of strangeness
production, i.e. $s_{\mathrm{th}} = (m_{D^*}+m_{\Lambda_c})^2 = 18.4 
\, \mathrm{GeV}^2$. 
In contrast with the $K^*\Lambda$ production, the effect of $D$
exchange is very much suppressed in the $D^* \Lambda_c$ production,
while $D^*$ exchange dominates the process, as shown in the left panel
of Fig.~\ref{FIG6}.  As mentioned in the case of the strangeness
production, the total cross section for the $\pi N\to D^*\Lambda_c$
reaction is proportional to $s^{J-1}$ when $s$ is large, so that 
$D^*$ exchange dictates the total cross section at higher energies. 
All other contributions including $D$ exchange have some effects on it
only in the vicinity of threshold.
The result of the total cross section for the $\pi^- p \to
D^{*-} \Lambda_c^+$  reaction is compared with that for the $\pi^- p
\to K^{*0}\Lambda$ in the right panel of Fig.~\ref{FIG6}. 
The total cross section for the charm production is about $10^4$ 
times smaller than that for the strangeness one near the threshold region 
and is about $10^3$ times smaller at around $s/s_{\mathrm{th}} = 10$.
It turns out that, in the effective Lagrangian method, this
suppression is mostly caused by the effect of form factors. 

\begin{figure}[thp]
\centering
\includegraphics[scale=0.3]{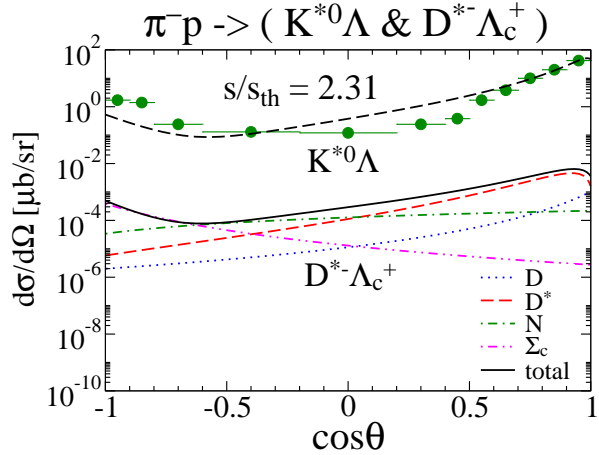}
\caption{Comparison of the differential cross section for the 
$\pi^- p \to D^{*-}\Lambda_c^+$ with that for the $\pi^- p \to K^{*0}\Lambda$
based on an effective Lagrangian approach.
The experimental data are taken from Ref.~\cite{Crennell:1972km}.}
\label{FIG7}
\end{figure}
The difference in the differential cross section $d\sigma/d\Omega$ is also
analyzed in Fig~\ref{FIG7}. As expected from the result of the total
cross section, $D^*$ exchange is dominant, particularly in the range
of $0\,\le \cos\theta \le 1$. In the backward region, $\Sigma_c$
exchange governs the charm process. 

\section{Regge Approach}
\begin{figure}[h]
\begin{tabular}{cc} 
\includegraphics[scale=0.5]{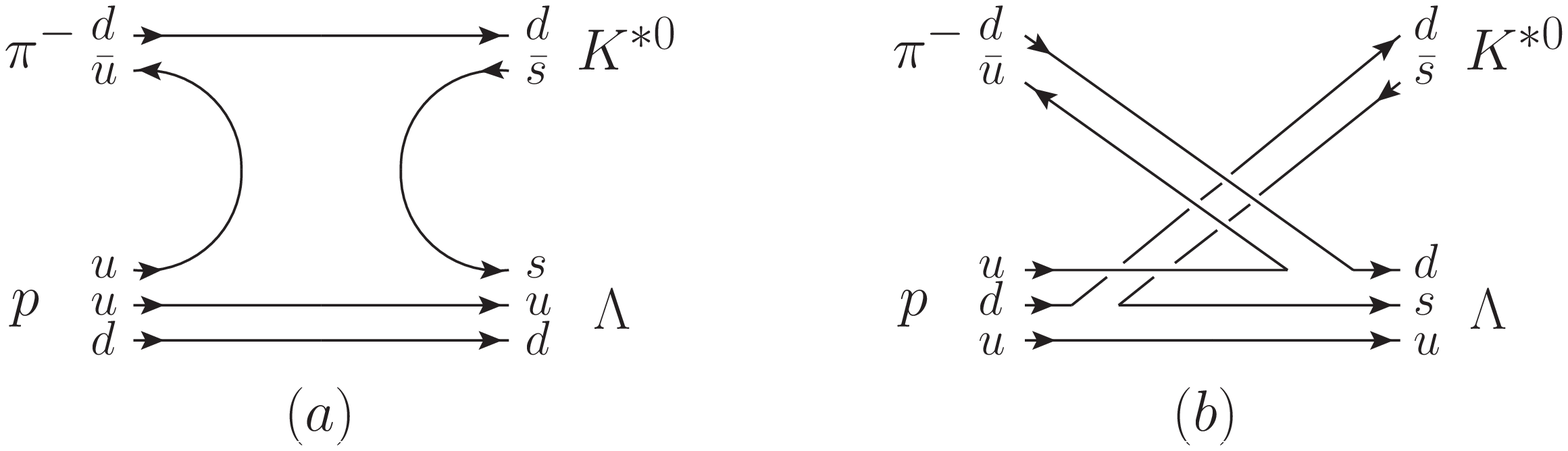}
\end{tabular}
\caption{Planar and nonplanar diagrams for the $\pi^- p \to
  K^{*0}\Lambda$ process in the left and right panels, respectively.}
\label{FIG8}
\end{figure}

Spurred on by the finding that the effective Lagrangian method describes
experimental data mainly in lower-energy regions in the previous
section, we will introduce in this section a Regge approach, which is
known to explain very well high-energy scattering with unitarity
preserved. The relevant diagrams for the strangeness production can be
schematically depicted in Fig.~\ref{FIG8} by the quark lines. There
are two different classes of diagrams: a planar diagram
[Fig.~\ref{FIG8}(a)] and a nonplanar diagram [Fig.~\ref{FIG8}(b)].  
In the Regge theory, the planar diagram is described by reggeon
exchange in the $t$-channel, whereas the nonplanar one corresponds to 
reggeon exchange in the $u$-channel.

\subsection{Regge amplitudes}
We first consider the $\pi^- p \to K^{*0}\Lambda$ reaction. 
The reggeons in the $t$-channel are dictated by the $K$ and $K^*$
trajectories, while the $\Sigma$-baryon trajectory leads to the reggeon 
in the $u$-channel, as displayed in Fig.~\ref{FIG8}.
In the present Regge approach, the Regge amplitudes are derived by  
replacing the Feynman propagator $P^{\mathrm{F}}$ contained in 
Eq.~(\ref{eq:EachAmpl}) by the Regge propagator
$P^{\mathrm{R}}$~\cite{Donachie2002},
\begin{eqnarray}
P_K^{\mathrm{R}}(s,t) &=& {1 \choose e^{-i\pi\alpha_K(t)}}
\left( \frac{s}{s_K} \right)^{\alpha_K(t)} 
\Gamma[-\alpha_K(t)] \alpha'_K,                                      \cr   
P_{K^*}^{\mathrm{R}}(s,t) &=& {1 \choose e^{-i\pi\alpha_{K^*}(t)}}
\left( \frac{s}{s_{K^*}} \right)^{\alpha_{K^*}(t)-1} 
\Gamma[1-\alpha_{K^*}(t)] \alpha'_{K^*},                               \cr
P_\Sigma^{\mathrm{R}}(s,u) &=& {1 \choose e^{-i\pi\alpha_\Sigma(u)}}
\left( \frac{s}{s_\Sigma} \right)^{\alpha_\Sigma(u)-\frac{1}{2}}
\Gamma\left[\frac{1}{2}-\alpha_\Sigma(u)\right] \alpha'_\Sigma,
\label{eq:RProp}
\end{eqnarray}
where $\alpha_{K}(t)$, $\alpha_{K^*}(t)$, and $\alpha_\Sigma(u)$ denote the 
Regge trajectories for the $K$ and $K^*$ mesons, and the $\Sigma$ baryon,
respectively. 
$s_K$, $s_{K^*}$, and $s_\Sigma$ stand for the energy scale parameters for 
the corresponding reggeons.
Thus the Regge amplitudes are represented by
\begin{eqnarray}
T_K(s,t) &=&
\mathcal{M}_K(s,t) 
P_K^{\mathrm{R}}(s,t) / P_K^{\mathrm{F}}(t),                                 \cr   
T_{K^*}(s,t) &=&
\mathcal{M}_{K^*}(s,t) 
P_{K^*}^{\mathrm{R}}(s,t) / P_{K^*}^{\mathrm{F}}(t),                           \cr
T_\Sigma(s,u) &=& 
\mathcal{M}_\Sigma(s,u) 
P_\Sigma^{\mathrm{R}}(s,u)  / P_\Sigma^{\mathrm{F}}(u),
\label{eq:RAmpl}
\end{eqnarray}
where $\mathcal{M}_K,\,\mathcal{M}_{K^*}$ and $\mathcal{M}_\Sigma$ are the
invariant Feynman amplitudes for the $K,\,K^*$, and $\Sigma$ exchanges,
respectively, as in Eq.~(\ref{eq:EachAmpl}) (and form factors are not 
included here).

\begin{figure}[h]
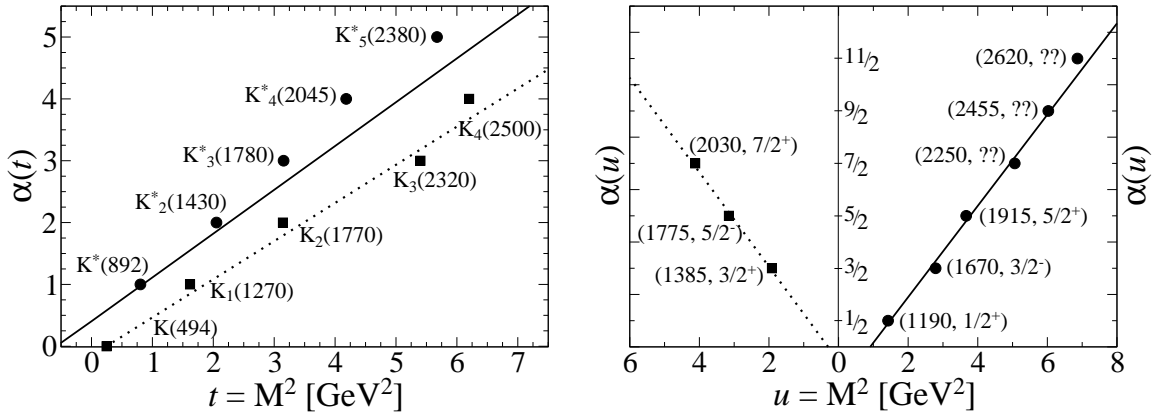

\vspace{1.5em}
\begin{tabular}{cc} 
\includegraphics[scale=0.3]{FIG9a.eps} \,\,\,\,\,
\includegraphics[scale=0.3]{FIG9b.eps}
\end{tabular}
\caption{$K$ and $K^*$ meson trajectories~\cite{Brisudova:2000ut} 
(left panel) and $\Sigma$ and 
$\Sigma^*$ trajectories~\cite{Storrow:1984ct} (right panel).}
\label{FIG9}
\end{figure}

The Regge trajectories for $K$ and $K^*$ are taken from
Ref.~\cite{Brisudova:2000ut}, respectively, as
$\alpha_K(t)=-0.151+0.617t$, $\alpha_{K^*}(t)=0.414+0.707t$.
The energy scale parameters are determined by using the
QGSM~\cite{Kaidalov:1982bq,Boreskov:1983bu,Kaidalov:1986zs,
Kaidalov:1994mda}: $s_K=1.752$ and $s_{K^*}=1.662$.
In general, a Regge propagator is expressed in terms of a linear
combination of the two different signatures. However, when a Regge
trajectory for a hadron with even spins is approximately the same as
that for a hadron with odd spins, that is, the two trajectories are almost
degenerate, one of the signatures is canceled out. As displayed in the 
left panel of Fig.~\ref{FIG9}, which is taken from
Ref.~\cite{Brisudova:2000ut}, both the $K$ trajectory and the $K^*$ one 
are almost degenerate. Thus the Regge propagator can have 
either the signature $1$ or $e^{-i\pi \alpha_K(K^*)}$ as shown in
Eq.~(\ref{eq:RProp})~\cite{Guidal:1997hy,Corthals:2005ce}. 

As for the $\Sigma$ trajectory, it is not easy to find some tendency
like the $K$ and $K^*$ trajectories. 
In the right panel of Fig.~\ref{FIG9}, we depict two trajectories for 
$\Sigma$s, assuming that the quantum  numbers for some unknown
resonances are fixed~\cite{Storrow:1984ct}. In the present
calculation, the solid trajectory is taken into account, for which
$\alpha_\Sigma(u)=-0.79+0.87u$~\cite{Storrow:1984ct}, since it 
contains the lowest-lying $\Sigma(1190)$. Based on this trajectory, we
are able to determine the scale parameter to be $s_\Sigma=1.569$ by using
the QSGM. 
We also assume that the $\Sigma$ trajectory is degenerated and two
different signatures 1 and  $e^{-i\pi\alpha_\Sigma(u)}$ are considered
as we did in the mesonic cases.

Since two different signatures are possible for each of reggeon 
exchanges, there are eight different ways of selecting the signatures for 
the $K$, $K^*$, and $\Sigma$ Regge propagators.
We have examined all the cases and have found that only the low-energy  
region ($1 \le s/s_{\mathrm{th}} \le 2$) is affected by the change in the 
phases by about 20$\%$. In the present calculation, we choose the
signature factor $1$ in common for all the Regge propagators.

A unique feature of the Regge amplitudes is that they can reproduce the
diffractive pattern both at forward and backward scatterings as well as 
the asymptotic behavior consistently with the unitarity.
Within the framework of a Regge approach, the differential cross
sections, $d\sigma/dt$ and $d\sigma/du$, must comply with the following 
forms asymptotically
\begin{eqnarray}
\frac{d\sigma}{dt}(s \to \infty, t \to 0) \propto s^{2\alpha(t)-2},\,\,\,
\frac{d\sigma}{du}(s \to \infty, u \to 0) \propto s^{2\alpha(u)-2}.
\label{eq:dsdt:dsdu}
\end{eqnarray}
Moreover, the present Regge amplitudes interpolate the low-energy
behavior near the threshold region and the high-energy (asymptotic)
behavior. 

\subsection{Normalization of Regge amplitudes}
In many high-energy processes, the absolute values of cross sections are 
determined empirically. For the effective Lagrangian method, this can
be done by employing a form factor. In the case of the Regge approach,
this may be done by considering an overall normalization factor.
In addition, some residual $t$ dependence is also included.
To fix the undetermined parameters, let us first examine the $s$ 
dependence of the total cross section and then the $t$ dependence of the 
differential cross section when taking into account each of the reggeon
exchanges separately in comparison with the experimental data for the 
$K^*\Lambda$ production.

\begin{figure}[thp]
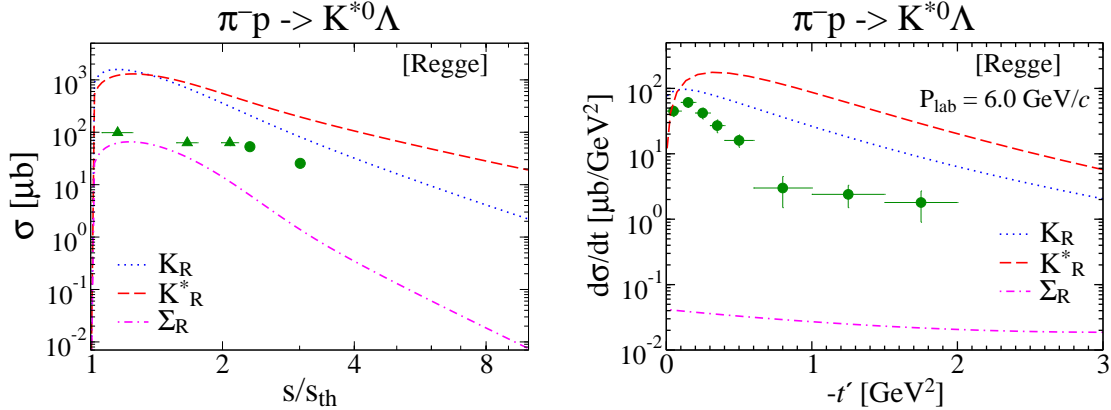

\begin{tabular}{cc} 
\includegraphics[scale=0.27]{FIG10a.eps} \;\;\;\;
\includegraphics[scale=0.27]{FIG10b.eps}
\end{tabular}
\caption{(Color online) Total cross sections for the 
$\pi^- p \to K^{*0} \Lambda$ based on the Regge approach without form 
factors. The data are taken from Ref.~\cite{Dahl:1967pg} (triangles) 
and Ref.~\cite{Crennell:1972km} (circles).}       
\label{FIG10}
\end{figure}
In the left panel of Fig.~\ref{FIG10}, the total cross section given
in Eq.~(\ref{eq:RAmpl}) is shown, with each reggeon contribution
separately drawn. To start with, let us take a look at the energy
dependence while the absolute values will be fixed later.
We observe that the $K^*$ reggeon term is in better agreement with the
data. The contributions of the $K$ and $\Sigma$ reggeons fall off faster
than the $K^*$ one, because of their smaller values of the intercept
$\alpha(0)$. This implies that the $K^*$ reggeon may play a dominant
role among the three reggeon contributions. 
Shown in the right panel of Fig.~\ref{FIG10} is the differential cross
section $d\sigma/dt$ as a function of $-t'$ at $\mathrm{P_{lab}} =
6.0\, \mathrm{GeV}/c$. At first glance, as $-t'$ is increased, both
the $K^*$ and $K$ reggeon terms seem to fall off more slowly than the
data. However, if we look at the small $|t'|$ region, to reproduce the  
sharp decrease at the forward angle, the $K^*$ reggeon
seems more important, though some contribution of the $K$ reggeon is
also required. 

To improve the $s$ and $t$ dependence simultaneously, we introduce an 
additional factor
\begin{eqnarray}
C_{\mathrm{ex}}(p^2) = \frac{a}{(1 - p^2/\Lambda^2)^2},
\label{eq:FF2}
\end{eqnarray}
which reflects a finite hadron size. Here $p$ stands for the transfer
momentum of the exchanged particle. The parameters $a$ and $\Lambda$
denote a dimensionless constant and a cutoff mass in units of GeV,
respectively. The parameter $a$ is introduced to fit the magnitude of
the amplitude. This residual factor $C_{\mathrm{ex}}(p^2)$ plays the
role of the form factor we have introduced in the effective Lagrangian
method. As will be shown in the next section, it greatly improves the $t$  
dependence of the differential cross section $d\sigma/dt$. Finally, 
we express the total result for the invariant Regge amplitude as
\begin{eqnarray}
T (\pi^- p \to K^{*0}\Lambda) =
T_K \cdot C_K + T_{K^*} \cdot C_{K^*} + T_\Sigma \cdot C_\Sigma.
\label{eq:SumAlpl}
\end{eqnarray} 

In the literature~\cite{Titov:2008yf}, a normalization factor 
$\mathcal{N}(s,t)$ has been introduced to reproduce the large $s$ behavior
by removing the extra $s$ and $t$ dependence possibly coming from the
interaction Lagrangian. The normalization factor 
is defined by  
\begin{eqnarray}
\mathcal{N}(s,t) = \frac{A^\infty(s)}{A(s,t)}, \,\,\,
A^2(s,t)= \sum_{s_i,s_f,\lambda_f}|{\mathcal{M}}(s,t)|^2,
\label{eq:NorFac}
\end{eqnarray}
where $A^\infty(s)$ is the dominant term when $s\to \infty$.
In the present case, however, such a factor is not needed, because the
amplitude (\ref{eq:SumAlpl}) already satisfies the desired large
$s$-behavior, and moreover the normalization factor (\ref{eq:NorFac})
removes the favorable $t$ dependence of the differential cross section
in the small $-t$ region.  In fact, the decreasing behavior of
$d\sigma/dt$ for small $-t$ arises from the $t$-dependent structure of
the effective Lagrangian amplitude that is incorporated in the 
amplitudes (\ref{eq:RAmpl}) and (\ref{eq:SumAlpl}). 

We can derive the Regge amplitudes for the charm production in a
similar way. Replacing the $s$ quarks in Fig.~\ref{FIG7} with $c$
quarks, we can draw the quark diagrams for the $\pi^- p\to D^{*-}
\Lambda_c^+$ process similar to Fig.~\ref{FIG7}.
The relevant amplitudes are written as  
\begin{eqnarray}
T_D(s,t) &=& \mathcal{M}_D(s,t) 
\left( \frac{s}{s_D} \right)^{\alpha_D(t)}
\Gamma\left[-\alpha_D(t)\right]   
\frac{\alpha'_D}{P_D^{\mathrm{F}}(t)},                                   \cr
T_{D^*}(s,t) &=& \mathcal{M}_{D^*}(s,t) 
\left( \frac{s}{s_{D^*}} \right)^{\alpha_{D^*}(t)-1}
\Gamma\left[1-\alpha_{D^*}(t)\right]   
\frac{\alpha'_{D^*}}{P_{D^*}^{\mathrm{F}}(t)},                             \cr  
T_{\Sigma_c}(s,u) &=& \mathcal{M}_{\Sigma_c}(s,u) 
\left( \frac{s}{s_{\Sigma_c}} \right)^{\alpha_{\Sigma_c}(u)-\frac{1}{2}}
\Gamma\left[\frac{1}{2}-\alpha_{\Sigma_c}(u)\right] 
\frac{\alpha'_{\Sigma_c}}{P_{\Sigma_c}^{\mathrm{F}}(u)}.
\label{eq:RAmpl:c}
\end{eqnarray}

\subsection{Results for $K^{*0}\Lambda$ production}
Having established the strategy above, we fix the strengths of the
free parameters $a$ and $\Lambda$ in $C_{\mathrm{ex}}(p^2)$ in 
Eq.~(\ref{eq:FF2}) by the following procedures:
\begin{itemize}
\item
The cutoff masses $\Lambda$ are chosen to be the typical
values:
$\Lambda_{K,\,K^*,\,\Sigma} = 1.0\,\mathrm{GeV}$.
\item
The $K^*$ reggeon dominance being known, its strength is determined by
the global $s$ and $t$ dependence of the observed $K^*\Lambda$
production cross sections: $a_{K^*} = 0.8$.
\item
The strength of the $K$ reggeon amplitude is chosen to reproduce the 
small $|t|$ behavior of $d\sigma/dt$ together with the dominant $K^*$ 
reggeon contribution: $a_K = 0.6$.
\item
The $\Sigma$ reggeon is determined to reproduce the backward peak 
behavior: $a_\Sigma = 1.5$.
\end{itemize}

\begin{figure}[thp]
\centering
\includegraphics[scale=0.5]{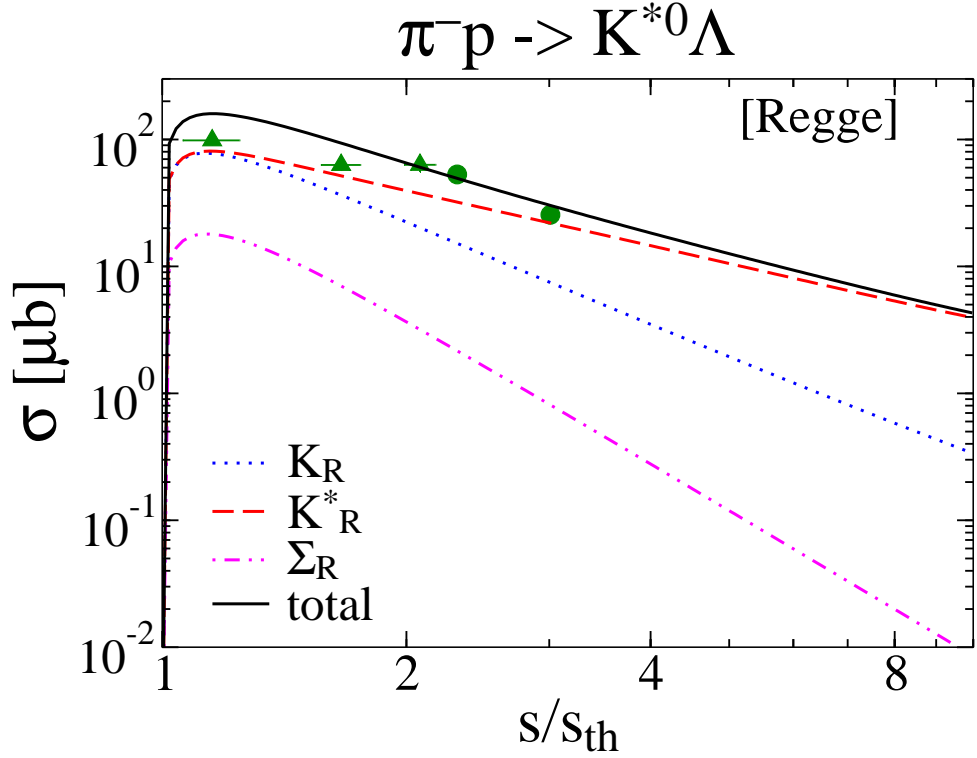}
\caption{(Color online).
Each contribution to the total cross sections
for the $\pi^- p \to K^{*0}\Lambda$ reaction given
as a function of $s/s_{\mathrm{th}}$,
based on a Regge approach.
The dotted and dashed curves show the contributions of 
$K$ reggeon exchange and $K^*$ reggeon exchange, respectively.
The dot-dot-dashed one draws the effect of $\Sigma$ reggeon exchange.
The solid curve represents the total result. 
The experimental data are taken from Ref.~\cite{Dahl:1967pg} 
(triangles) and from Ref.~\cite{Crennell:1972km} (circles).}         
\label{FIG11}
\end{figure}
In Fig.~\ref{FIG11}, the total cross section is illustrated, together with 
each contribution. $K^*$ reggeon exchange governs its 
dependence on $s$. The contribution of $K$ reggeon exchange is smaller
than that of $K^*$ reggeon exchange, which becomes clear as $s$
increases. The reason is obvious from the value of $\alpha_K(t)$
mentioned previously: the corresponding intercept is smaller than that
for the $K^*$ trajectory and its slope is steeper than that of the $K^*$ 
one.
We have seen in Fig.~\ref{FIG3} that the contribution of $K^*$
exchange in the effective Lagrangian method rises slowly as $s$
increases, which results in deviation from the experimental data. On
the other hand, $K^*$ reggeon exchange exhibits the $s$ dependence of
the total cross section correctly, so that it describes the experimental 
data much better than $K^*$ exchange in the effective Lagrangian method at 
higher values of $s$. 
It is interesting to see that $\Sigma$ reggeon exchange in the
$u$-channel contributes to the total cross section approximately by
$20\,\%$ in the vicinity of threshold whereas its effect becomes much
smaller as $s$ increases. This can be understood from the behavior of
the $u$-channel Regge amplitude:$T_{\Sigma}\sim s^{-0.79}$. Note that
this feature of $\Sigma$ reggeon exchange is significantly different
from that of $\Sigma$ exchange in the effective Lagrangian method,
where the $u$-channel makes a negligibly small contribution (see 
Fig.~\ref{FIG3} for comparison).

\begin{figure}[thp]
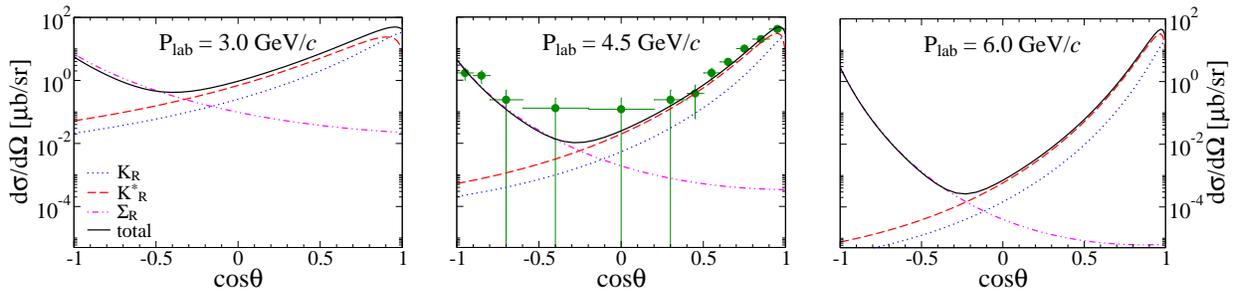

\centering
\includegraphics[width=5.30cm]{FIG12a.eps}\hspace{1.1em}
\includegraphics[width=4.55cm]{FIG12b.eps}\hspace{1.1em}
\includegraphics[width=5.30cm]{FIG12c.eps}
\caption{(Color online). 
Differential cross sections for the $\pi^- p \to K^{*0}\Lambda$ reaction 
as functions of $\cos\theta$ 
at three different pion momenta ($\mathrm{P_{lab}}$), 
based on a Regge approach. 
The experimental data are taken from Ref.~\cite{Crennell:1972km}.
The notations are the same as Fig.~\ref{FIG11}.}
\label{FIG12}
\end{figure}
Figure~\ref{FIG12} depicts the results of the differential cross
section $d\sigma/d\Omega$ for the $\pi^- p \to K^{*0}\Lambda$
reaction. The $K^*$ reggeon in the $t$-channel makes a
dominant contribution to the differential cross section in the forward
region, whereas the $\Sigma$ reggeon in the $u$-channel enhances
it at the backward angles. The effect of $K$ reggeon exchange is 
important to describe the experimental data at the very forward
angle. We already have found that the results from 
the effective Lagrangian method deviate from the experimental data  
except for the forward region. This is to a great extent due to the
fact that the $u$-channel contribution is underestimated in the
effective Lagrangian method. However, the Regge approach correctly
describes the experimental data at
$\mathrm{P_{lab}}=4.5\,\mathrm{GeV}/c$  
over the entire angles. Moreover, on the whole, it elucidates the
flatness of the differential cross section between $\cos\theta=0.3$
and $\cos\theta=-0.7$, which was never explained in the effective
Lagrangian method.

\begin{figure}[htp]
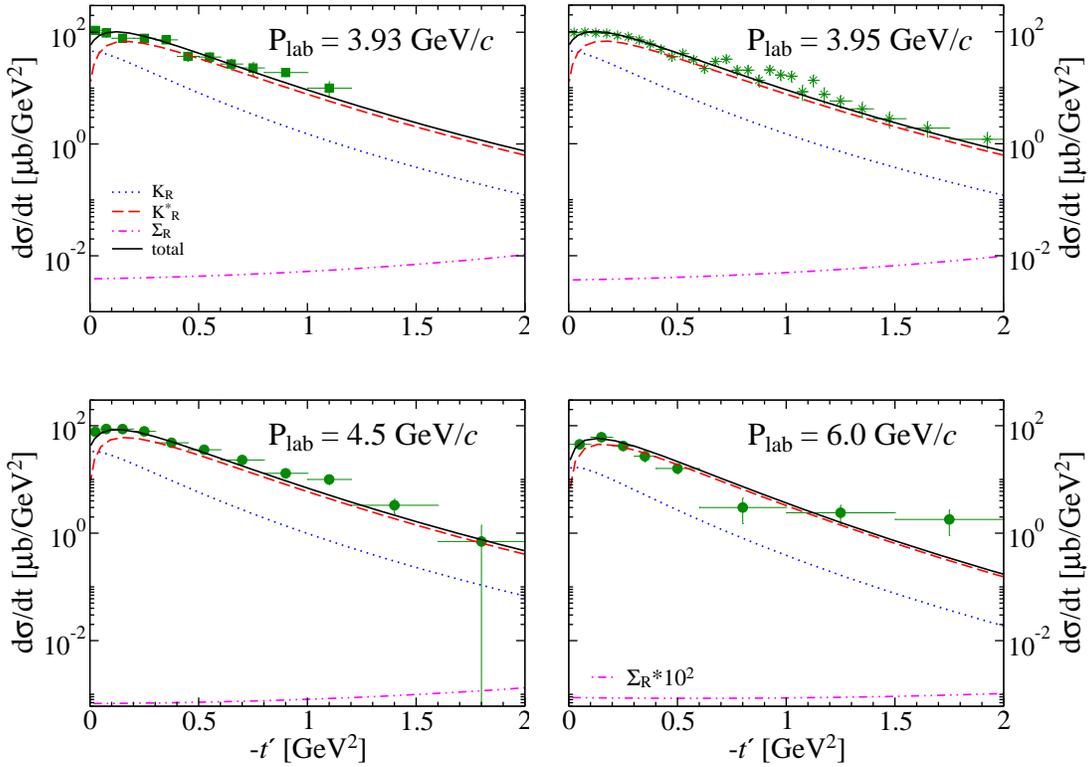

\vspace{1em}
\begin{tabular}{cc}
\includegraphics[width=7.0cm]{FIG13a.eps}\,\,\,\,
\includegraphics[width=7.0cm]{FIG13b.eps}\vspace{1.73em} \\ 
\includegraphics[width=7.0cm]{FIG13c.eps}\,\,\,\,
\includegraphics[width=7.0cm]{FIG13d.eps}
\end{tabular}
\caption{(Color online).
Differential cross sections for the $\pi^- p \to K^{*0}\Lambda$ reaction 
as functions of $-t^\prime$ at four different pion momenta
($\mathrm{P_{lab}}$), based on a Regge approach. 
The experimental data denoted by the squares are taken from 
Ref.~\cite{Yaffe:1973ex}, 
while those denoted by the stars are from 
Ref.~\cite{Aguilar-Benitez:1980pd}. 
Those designated by the circles are taken from 
Ref.~\cite{Crennell:1972km}. 
The notations are the same as Fig.~\ref{FIG11}.}
\label{FIG13}
\end{figure}
In Fig.~\ref{FIG13}, we draw the results of the 
$\pi^- p \to K^{*0}\Lambda$ differential cross section $d\sigma/dt$ as
functions of $-t'$ at four different values of $\mathrm{P_{lab}}$.
The most dominant contribution comes from $K^*$ reggeon exchange. 
$K$ reggeon exchange plays a crucial role in explaining the 
data at the very forward angle together with $K^*$ reggeon exchange. A
similar feature can also be found in the case of $K\Lambda$
photoproduction~\cite{Guidal:1997hy}.  The effect of $\Sigma$ reggeon 
exchange turns out to be tiny.  Though the general feature of the 
results from the Regge approach looks apparently similar to that of the 
effective Lagrangian ones, they are in fact different from each other. 
The results from the Regge approach fall off faster than those from the 
effective Lagrangian method, as $-t'$ increases. As a result, the
Regge approach reproduces the experimental data better in 
comparison with the effective Lagrangian method.  

\subsection{Results for $D^{*-}\Lambda_c^+$  production}
\begin{figure}[htp]
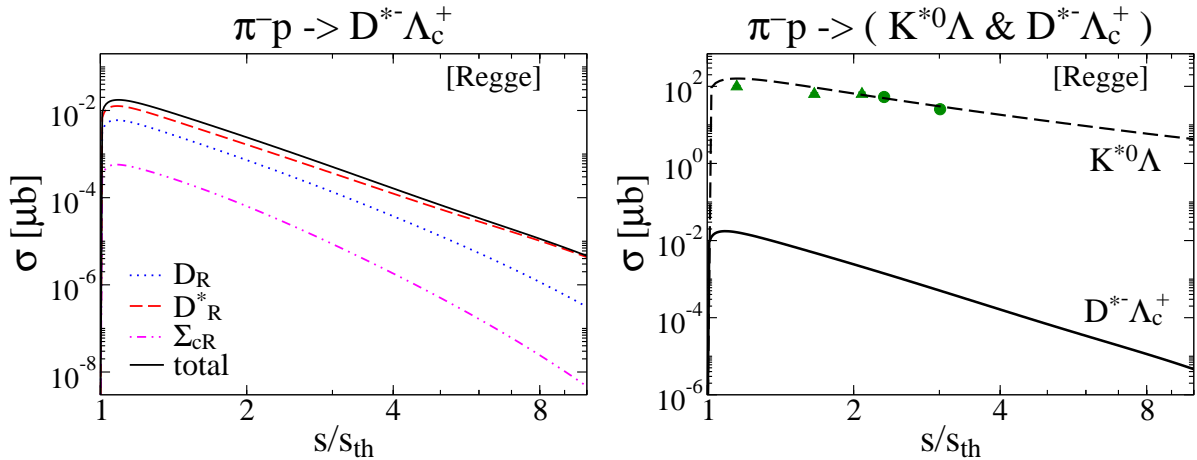

\begin{tabular}{cc} 
\includegraphics[scale=0.3]{FIG14a.eps} \;
\includegraphics[scale=0.3]{FIG14b.eps}
\end{tabular}
\caption{(Color online).
In the left panel, each contribution to the total cross sections
for the $\pi^- p \to D^{*-}\Lambda_c^+$ reaction is drawn 
as a function of $s/s_{\mathrm{th}}$ 
from a Regge approach. The dotted and dashed curves show $t$-channel
contributions, i.e. those of $D$ reggeon exchange and $D^*$ reggeon
exchange, respectively. The dot-dot-dashed curve depicts the
contribution of $\Sigma_c$ reggeon exchange. 
The solid curve represents the full result of the total cross section. 
In the right panel, the total cross section for 
the $\pi^- p \to D^{*-}\Lambda_c^+$ reaction (solid curve) is compared with
that for the $\pi^- p \to K^{*0}\Lambda$ one (dashed one).
The experimental data for the $\pi^- p \to K^{*0}\Lambda$ reaction are
taken from Ref.~\cite{Dahl:1967pg} (triangles) and
from Ref.~\cite{Crennell:1972km} (circles).} 
\label{FIG14}
\end{figure}

We now discuss the results of the charm production. 
In the left panel of Fig.~\ref{FIG14}, we draw the total cross section
together with each contribution for the $\pi^- p \to D^{*-}\Lambda_c^+$
reaction. $D^*$ reggeon exchange dictates the $s$ dependence of the
total cross section. The contributions of $K$ reggeon and $\Sigma_c$ 
reggeon exchanges are more suppressed than that of $K^*$ reggeon
exchange. In the right panel of Fig.~\ref{FIG14}, we compare the
$D^*\Lambda_c$ production with the $K^*\Lambda$ one.
It is found that the total cross section for the charm production is 
approximately $10^4 - 10^6$ times smaller than that for the strangeness 
production depending on the energy range of $s/s_{\mathrm{th}}$.
The resulting production rate for $D^*\Lambda_c$ at $s/s_{\mathrm{th}}
\sim 2$, which is the energy that can excite charmed baryons up to
$\sim$ 1 GeV,
is suppressed by about a factor of $10^4$ in comparison with
the strangeness production. This implies that the production cross
section of $D^*\Lambda_c$ is around 2 $\mathrm{nb}$ at that energy. 

In fact, one of the present authors carried out a similar study
~\cite{Noumi:2014vfa} based on a Regge method from 
Ref.~\cite{Grishina:2005cy}, where a phenomenological form factor was 
included in the Regge expression for the total cross section.  
As illustrated in Fig.~3 in Ref.~\cite{Noumi:2014vfa}, the total cross
section for  the $D^*\Lambda_c$ production was shown to be
approximately $10^4$ times smaller than that for the $K^* \Lambda$
production at $s/s_{\mathrm{th}} \sim 2$, 
which is of almost the same order compared with the present
result. However, one has to keep in mind that the form factor introduced
by Ref.~\cite{Grishina:2005cy} bears no relation to the effective 
Lagrangian method. This is understood from the observation that near
the threshold the result of this model is too much underestimated
compared with that of the effective Lagrangian method.

\begin{figure}[ht]
\centering 
\includegraphics[scale=0.5]{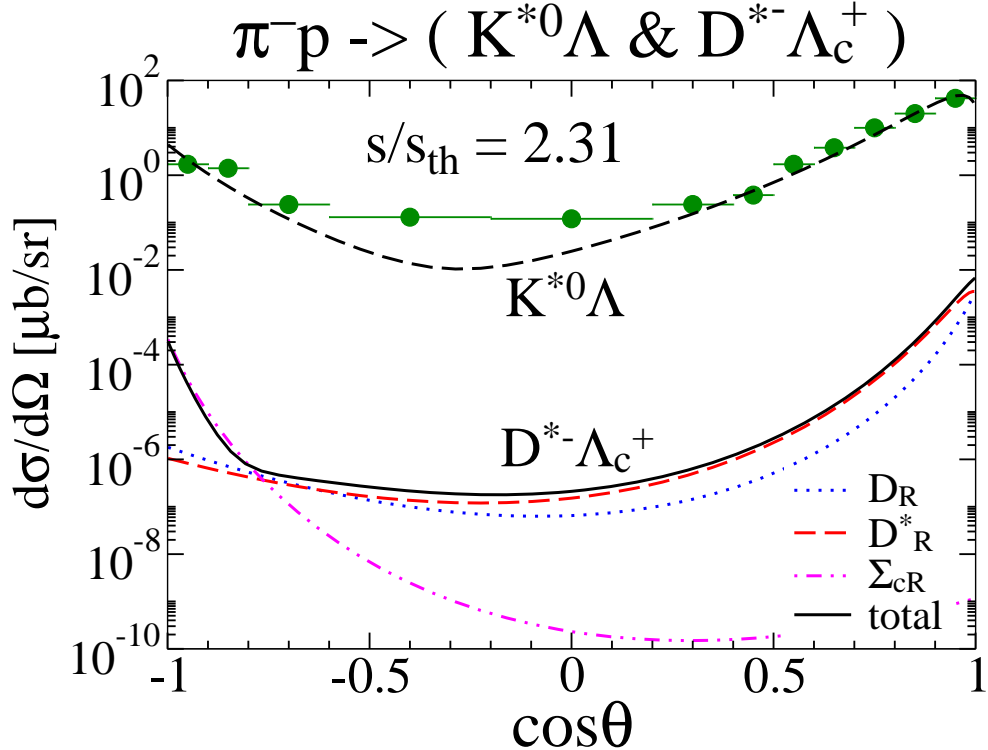}
\caption{Comparison of the differential cross section for the 
$\pi^- p \to D^{*-}\Lambda_c^+$ with that for the $\pi^- p \to K^{*0}\Lambda$
based on a Regge approach.
The experimental data are taken from Ref.~\cite{Crennell:1972km}.} 
\label{FIG15}
\end{figure}
In Fig.~\ref{FIG15}, the results of the differential cross section
$d\sigma/d\Omega$ for both strangeness and charm productions are also
compared to each other. $D^*$ reggeon exchange plays a crucial role
throughout the whole angle region. However, its contribution is 
diminished as $\cos\theta$ decreases, compared with effective
Lagrangian method (see Fig.~\ref{FIG7}).

\section{Comparison of the two models}
To analyze what causes the large difference in the cross sections
between the strange and charmed productions, we first calculated 
the cross section without considering form factors.
In the effective Lagrangian method, it is interesting that, when 
excluding the Feynman propagators in Eq.~(\ref{eq:EachAmpl}) as well as 
the form factors, each contribution for the charm production is even 
larger than that for the strangeness one within a factor of 10 except for 
the $N$ exchange.
In the case of $N$ exchange, the difference is between the factors of 10 
and 100.
Since the energy scale for the charm production is larger than the
strangeness one, the Feynman propagator suppresses the charm production 
much more than the strangeness one.
The form factor also contains the $(p^2 -M_{ex}^2)$ term in the denominator
which suppresses the amplitude more because it is a second power.
However, in the case of the Regge approach, the
result of the cross section when excluding form factors is quite
different from that of the effective Lagrangian method.
The form factors barely affect the difference in the cross sections.

Considering the fact that Ref.~\cite{Christenson:1985ms} has 
experimentally measured only an upper limit $\sigma \sim 7\,\mathrm{nb}$ 
at the pion momentum $\mathrm{P_{lab}} = 13 \,\mathrm{GeV}/c$ for the 
charm production, we find that the results derived from our models
are within a factor of 2 from this upper limit: 
$\sigma = 14\,(9)\,\mathrm{nb}$
when employing the effective Lagrangian method (Regge approach).
However, some ambiguity lies in the selection of cutoff masses.
If we apply slightly smaller cutoff masses, for example, $0.5\, (0.9)$
GeV for the effective Lagrangian method (Regge approach), our results
will underestimate the upper limit 7 nb without influencing the general
results for the strangeness productions.
The slope of the differential cross section also mildly changes with
the cutoff masses varied. 

\section{Summary and Conclusion}
In the present work, we aimed at describing both the strangeness and
charm productions by the pion beam, based on both an effective
Lagrangian method and a Regge approach. We started with the effective
Lagrangian method to describe the $\pi N \to K^*\Lambda$ and 
$\pi N \to D^*\Lambda_c$ reactions. The coupling constants were
determined either by using the experimental data or by employing those
from a nucleon-nucleon potential and flavor SU(3) symmetry. 
The cutoff masses of the form factors were fixed to reproduce the 
experimental data. However, in order to reduce the ambiguity in the 
effective Lagrangian method, we used the equal values of the cutoff 
masses for each case of meson exchange and baryon exchange.

We were able to explain the total cross section for the 
$\pi^- p \to K^{*0}\Lambda$ in lower-energy regions within the framework of
the effective Lagrangian method. However, the results from the effective
Lagrangian method start to deviate from the data, as the square of the
total energy $s$ increases. The magnitude of the total cross section for 
the $\pi^- p \to D^{*-}\Lambda_c^+$ was approximately $10^3$ times
smaller than that for the $\pi N\to K^*\Lambda$. The  
$t$-channel contributes to the differential cross section in the forward
direction, whereas the $u$-channel contributes to that in the backward
direction. The differential cross section $d\sigma/dt$ for the
$\pi^-p\to K^{*0}\Lambda$ tends to decrease as $-t'$ increases. The
results of $d\sigma/dt$ were in good agreement with the experimental
data at lower $\mathrm{P_{lab}}$. 

We constructed the Regge propagators for $K$ and $K^*$ reggeons. Since 
the corresponding trajectories are degenerate, we were able to
consider the signature either to be 1 or to be a complex phase. 
The $\Sigma$ reggeon was also considered for the description of the 
backward angle region. We selected 1 as the signatures for all the
reggeon propagators. Our Regge model satisfies the asymptotic
behavior of the total cross section to a great extent as $s$ becomes
very large. The difference between the strange and charmed total cross
sections turns out to be $10^4 - 10^6$, depending on the energy range. 
Compared with the results from the effective Lagrangian method, the Regge 
approach describes the experimental data better, in particular, in
higher-energy regions.  

In the present paper, our estimation corresponds to the production 
rates of the ground state $\Lambda_c$ associated with a charmed meson 
$D^*$.
On the other hand, production rates of various excited states together 
with their decays are related to their structures formed by the heavy- 
and light-quark contents.
The relevance of production rates to different structures of excited 
states has been addressed previously~\cite{Kim:2014qha}.
The identification of spin doublets, e.g., $J^P = 1/2^-$ and $3/2^-$ 
states, will clarify the nature of heavy-quark spin symmetry.
The identification of different internal modes, the so-called $\lambda$ 
and $\rho$ modes~\cite{Copley:1979wj}, can address how two light quarks 
(diquarks) are excited inside a baryon.
The excited diquark may couple to the pion to decay, carrying basic 
information of chiral symmetry.
In future experiments, we hope to see such fundamental issues of the
physics of the strong interaction.

We want to mention that $N^*$ resonances~\cite{Kim:2011rm,Kim:2014hha}
were not considered, because we are mainly interested in higher-energy
regions, where the experimental data are available to date. However,
it is of great interest to take into account $N^*$ resonances, when
one wants to understand the mechanism of the $K^*\Lambda$ production
near threshold in detail. In addition, since the $K^*$ meson in the
final state is a vector meson, it is very important to understand
polarization observables and density matrix elements. 
Furthermore, we can extend the present work to the reaction
$\pi^- p \to K^{*0}\Sigma^0 \, (K^{*+}\Sigma^-)$ and the corresponding charm
production $\pi^- p \to D^{*-} \Sigma_c^+ \, (\bar D^{*0} \Sigma_c^0)$.
The corresponding results will appear elsewhere soon.

\section*{Acknowledgments}
This work is supported in part by the Grant-in-Aid for Science Research 
(C) 26400273.  
S.H.K. is supported by a Scholarship from the Ministry of Education, 
Culture, Science and Technology of Japan.
The work of H.-Ch.K. was supported by the Basic Science Research Program 
through the National Research Foundation of Korea funded by the Ministry 
of Education, Science and Technology (Grant N0. 2013S1A2A2035612).

\end{document}